\documentclass[letterpaper,twocolumn,10pt]{article}
\usepackage{usenix,epsfig,endnotes,comment,subfigure, multirow, cite, color, amsfonts, pifont, algorithm2e}                  

\usepackage[hyphens]{url}
\usepackage{hyperref}
\usepackage{breakurl}

\begin{document}

%don't want date printed
\date{}

%make title bold and 14 pt font (Latex default is non-bold, 16 pt)
\title{\Large \bf CensorSpoofer: Asymmetric Communication with IP Spoofing for Censorship-Resistant Web Browsing}

%for single author (just remove % characters)

%\numberofauthors{1}

\author{
{\rm Qiyan Wang$\dagger$}\\	
\and
{\rm Xun Gong$\ddagger$}\\
% copy the following lines to add more authors
 \and
 {\rm Giang T. K. Nguyen$\dagger$}\\
 \and
 {\rm Amir Houmansadr$\ddagger$}	\\ 
 \and
 {\rm Nikita Borisov$\ddagger$} \vspace{2mm} \\ 
 $\dagger$Department of Computer Science  \\
 $\ddagger$Department of Electrical and Computer Engineering \\
 University of Illinois at Urbana-Champaign	\\
 {\{qwang26, xungong1, nguyen59, ahouman2, nikita\}@illinois.edu}
} % end author

\maketitle

% Use the following at camera-ready time to suppress page numbers.
% Comment it out when you first submit the paper for review.
\thispagestyle{empty}

\subsection*{Abstract}

A key challenge in censorship-resistant web browsing is being able to direct legitimate users to redirection proxies while preventing censors, posing as insiders, from discovering their addresses and blocking them.  We propose a new framework for censorship-resistant web browsing called {\it CensorSpoofer} that addresses this challenge by exploiting the asymmetric nature of web browsing traffic and making use of IP spoofing.  CensorSpoofer de-couples the upstream and downstream channels, using a low-bandwidth indirect channel for delivering outbound requests (URLs) and a high-bandwidth direct channel for downloading web content.  The upstream channel hides the request contents using steganographic encoding within email or instant messages, whereas the downstream channel uses IP address spoofing so that the real  address of the proxies is not revealed either to legitimate users or censors.  We built a proof-of-concept prototype that uses encrypted VoIP for this downstream channel and demonstrated the feasibility of using the CensorSpoofer framework in a realistic environment.

\section{Introduction}
\label{sec:introduction}

%First, show that the problem of censorship is very severe. 

Today, the Internet is playing an ever-increasing role in social and political movements around the world.  Activists use it to coordinate their activities and to inform the general people of important information that is not available via traditional media channels. The role played by Twitter, Facebook, YouTube, CNN~iReport and many other websites/blogs in the recent events in the Middle East is a great example of this~\cite{Egypt, Iran}.

The free flow of information and exchange of ideas on the Internet has been perceived as a serious threat by repressive regimes. In response, they have imposed strong censorship on the Internet usage of their citizens. They monitor, filter, trace, and block data flows using sophisticated technologies, such as IP address blocking, DNS hijacking, and deep packet inspection~\cite{gifc:07,Leberknight2010}. For example,  the ``Great Firewall of China'' blocks almost all popular social networks, such as Facebook, Twitter and Flickr, and other websites that may provide political information contrary to the state's agenda, such as Youtube, Wikipedia, BBC News, and CNN~\cite{zit03}.
To exercise control over the Internet, the Chinese government employs an Internet police force of over 30\,000 people to constantly monitor the citizens' online activities~\cite{internet_police}, and an individual who is caught violating the laws of Chinese censorship could be forced to pay a fine of up to \$1800 or sent to jail~\cite{punishment}.

There are many tools that aim to circumvent such censorship~\cite{dynaweb,ultrasurf,psiphon,tor}; a typical approach is to deploy a redirection proxy that provides access to blocked sites. Censors are, however, eager to locate such proxies and block them as well; a particularly powerful approach is the \emph{insider attack}, wherein censors pretend to be legitimate users of the service in order to locate and shut down the proxies.  Limiting the amount of information each user gets and trying to identify compromised insiders can partially mitigate this attack~\cite{mahdian10,sovran08,mccoy11}; however, these techniques are unlikely to survive a powerful adversary who can deploy a very large number of corrupt users.  An alternate approach is to never reveal the proxies' address to legitimate users and thus be completely immune to the insider attack.  Some recent work suggests strategically placing special deflection routers at core Internet ISPs to transparently redirect users' traffic to the proxies~\cite{Karlin2011, Wustrow2011,Houmansadr2011a}.  Such a deployment, however requires a significant resource investment that is likely to come only from a (pro-Internet freedom) government agency, as well as cooperation of large ISPs.

We propose a new approach, CensorSpoofer, that can be deployed using minimal resources, perhaps volunteered by ordinary people interested in promoting Internet freedom. (The Tor project~\cite{tor} has demonstrated the feasibility of building a successful service with contributions from such volunteers.)  Our key insight is that it is possible to use IP address spoofing to send data from the proxy to a user without revealing its actual origin.  Such a spoofed channel allows communication in a single direction only; however, we can exploit the asymmetric nature of web-browsing traffic, using a low-bandwidth indirect channel, such as steganographic instant messages or email, to communicate requests from the user to the proxy.  To avoid identification by the censor, CensorSpoofer mimics an encrypted VoIP session to tunnel the downstream data, since the VoIP protocol does not require endpoints to maintain close synchronization and does not reveal its contents to the censor.  We also explore additional steps that need to be taken to prevent detection; namely, choosing a plausible fake IP source address.

To demonstrate the feasibility of CensorSpoofer, we built a proof-of-concept prototype implementation and tested it in a real-world environment.  Our experiments show that our prototype can be successfully used for browsing the web while resisting blocking efforts of the censors. % XXX if there's a comparison with Tor, add it here.

%Fifth, roadmap.
The rest of this paper is organized as follows. We introduce the related work in Section~\ref{sec:related_work}. Section~\ref{sec:concept} presents the basic concepts, including the threat model and system goals. Section~\ref{sec:framework} describes the framework of CensorSpoofer. In Section~\ref{sec:design}, we elaborate a concrete design of CensorSpoofer based on VoIP, and analyze its security in Section~\ref{sec:security}. Section~\ref{sec:experiment} presents our prototype implementation and the evaluation results. We conclude in Section~\ref{sec:conclusion}.

\section{Related Work}
\label{sec:related_work}

\begin{comment}

\begin{table*}[t]
\centering
\caption{Comparison of circumvention systems that support web browsing.}
\label{tab:compare}

\begin{small}

\begin{tabular}{ c c c c c }
\hline
\multirow{2}{*}{}			&	Require special 	& Require support & Require unblockability 	& Resist to any number	\\
	&	servers & from ISPs	& of encrypted traffic	& of corrupted users	\\
\hline

Tor~\cite{tor}, Ultrasurf~\cite{ultrasurf}, etc.								&	\ding{55}		& \ding{55}	& \checkmark	& \ding{55}		\\

Infrastructure-assisted~\cite{Karlin2011, Wustrow2011,Houmansadr2011a}	
			&	\ding{55}		& \checkmark	& \checkmark	& \checkmark	\\
			
MailMyWeb~\cite{mailmyweb}, FOE~\cite{foe}											&	\checkmark		& \ding{55}	& \checkmark	& \checkmark	\\

%Collage~\cite{collage}		&	\ding{55}		& \ding{55}	& \ding{55}	& \checkmark	& \ding{55}	\\

CensorSpoofer	 &	\ding{55}		& \ding{55}	& \checkmark	& \checkmark	\\

\hline
\end{tabular}

\end{small}
\end{table*}

\end{comment}

%\subsection{Censorship Circumvention}
                                                   
In response to Internet censorship, many pragmatic systems  such as Dynaweb/freegate~\cite{dynaweb}, Ultrasurf~\cite{ultrasurf}, and Psiphon~\cite{psiphon} have been developed to help people bypass censorship. All these systems are based on a simple idea: let the user connect to one of the proxies deployed outside the censor's network, which can fetch blocked webpages for the user.  To hide the nature of the traffic, the communications with the proxy are encrypted.  Infranet~\cite{infranet} takes things a step further, embedding the real communication inside a cover web session, using covert channels to communicate the request and image steganography to return the data.  However, while escaping detection by outsiders, these designs are vulnerable to the insider attack, where the censor pretends to be an ordinary user to learn the location of the proxies and then block them.

Tor~\cite{tor} also uses proxies (called {\it bridges}, run by volunteers) to resist censorship, but employs more advanced strategies to limit the distribution of proxies' IP addresses.  So far, Tor has tried four different distribution strategies.  First, each user would receive a small subset of bridges based
on their IP address as well as the current time.  Second, a small subset could be obtained by sending a request via GMail.  These strategies fail to protect against an adversary who has access to a large number of IP addresses and GMail accounts; Chinese censors were able to enumerate all bridges in under a month~\cite{bridge-distr}.  (McLachlan and Hopper further showed that open proxies could be used to gain access to a large number of IP addresses~\cite{McLachlan09}).  The third strategy involves distributing bridge addresses to a few trusted people in censored countries in an ad hoc manner,
who then disseminate this information to their social networks.  Fourth, an individual can deploy a private bridge and give the bridge's address only to trusted
contacts.  These methods can resist bridge discovery but reach only a limited fraction of the population of potential bridge users.

Several researchers have tried to design better relay distribution strategies~\cite{mahdian10, sovran08, mccoy11, Feamster03} that aim to identify users who are likely to lead to a relay being blocked using past history and directing new relay information towards other users.  However, these designs are not likely to withstand a censor who controls a large number of corrupt users.

Another school of research on censorship circumvention tries to fundamentally resist the insider attack, i.e., tolerating any fraction of corrupted users. The idea is to hide the relay's IP from any user and therefore the censors. One way to achieve that is to utilize indirect channels, i.e., relaying the traffic sent to/by the relay through one or more intermediate nodes. For example, MailMyWeb~\cite{mailmyweb} and FOE~\cite{foe} utilize Email as the indirect channel. For these systems,  users are required to be able to access foreign servers that support encryption (e.g., Gmail), in order to avoid being detected by the censor. Nevertheless, considering the Chinese government once temporarily blocked Gmail~\cite{gmail_block}, we can envision the censor would again block the few special email providers, once finding out they are popularly used to bypass censorship.

It is important to note that, while an indirect channel is also used in CensorSpoofer, we only use it for sending outbound messages (e.g., URLs), which are usually very small (especially after encoding URLs into small numbers) and easy to hide into any indirect channel using steganography. This allows us to obviate the need for special servers (e.g., external Email providers supporting encryption) to provide a secured and high-bandwidth indirect channel. Consequently, the cost of blocking the outbound channel of CensorSpoofer is significantly higher: the censor has to block all overseas indirect communication (e.g., overseas Email and IM) even though the users only use the local Email and IM providers controlled by the censor.

More recently, researchers proposed several infrastructure-assisted circumvention systems, including Telex~\cite{Wustrow2011}, Decoy routing~\cite{Karlin2011}, and Cirripede~\cite{Houmansadr2011a}. Although these systems can support low-latency communication and perfectly resist the insider attack, they require a significant investment of effort by core Internet ISPs.  By contrast, CensorSpoofer is an infrastructure-independent circumvention system, allowing individuals to deploy their own anti-censorship systems without requiring any additional support from network infrastructure.

Instead of aiming to provide low-latency communication, some anti-censorship systems are designed to achieve censorship-resistant content sharing and/or distribution. For example, some works leverage peer-to-peer (P2P) networks to provide privacy-preserving file sharing, e.g., Freenet~\cite{freenet}, membership concealing overlay network~\cite{mcon}, and darknet~\cite{waste, turtle}. 
Collage~\cite{collage} let users stealthily exchange censored information with an external relay via a website that can host user-generated content (e.g., Flickr) using steganography.

\section{Concept}
\label{sec:concept}

\subsection{Threat Model}
\label{ssec:threat_model}

We consider a state-level adversary (i.e., the censor), who controls the network infrastructure under its jurisdiction. The censor has sophisticated capabilities of IP filtering, deep packet inspection, and DNS hijacking, and can potentially monitor, block, alter, and inject traffic anywhere within or on the boarder of its network.
However, the censor is motivated to allow citizens to {\it normally} access basic Internet services, such as IM, email and VoIP, as blocking such services would lead to economic lesses and political pressure.  
More specifically, we assume the censor is unwilling to interfere with the Internet connections of a user, e.g., an ongoing VoIP conversation, unless it has evidence that a particular connection is being used for bypassing censorship. 

Furthermore, we assume the censor generally allows people to use common encryption protocols to protect their online privacy, e.g., SRTP~\cite{srtp} for secure VoIP communication.  Thus far, this assumption has held true for most existing cases of Internet censorship, and the use of encrypted protocols such as SSL/TLS have formed the foundation of most existing anti-censorship  systems~\cite{tor, dynaweb, ultrasurf, psiphon, mailmyweb, foe, Wustrow2011, Karlin2011, Houmansadr2011a}. 
Once again, blocking encrypted traffic reduces the security of normal citizens using the Internet for personal or business reasons, and thus censors are motivated to allow such traffic through.  There have been important exceptions to this, including Iran's blocking of all encrypted traffic prior to the 33rd anniversary of the Islamic Revolution~\cite{block_https} and Egypt's complete disconnection of the Internet in response to nationwide protests~\cite{egypt-block-all}.  Such drastic censorship requires  fundamentally different circumvention approaches that are out of scope of our work.                      

%Nevertheless, we also note that, it is still possible that highly stringent censors may block all encrypted traffic regardless, to enforce the maximum restriction on Internet access. For example, the Iranian government temporarily blocked TLS/SSL before the 33rd anniversary of the Islamic Revolution, causing inaccessibility to all HTTPS websites such as secure Email and online banking sites~\cite{block_https}.  Under such strong attacks, all existing circumvention systems used for web browsing are unable to work. For clarity, such scenarios are outside our threat model; however, we assume the censor can block {\it any} entities on the Internet that generate encrypted traffic, such as any IM, Email, and VoIP servers that employ encryption, if it has reasons to believe they are used to circumvent censorship. 

We assume the censor can utilize its governmental power to force local IM, Email, and VoIP providers to censor their users' communication.  We also 
assume that the censor can block \emph{any} foreign Internet website or service, such as an email or instant messaging provider, if it has reason to believe
that it is being used to circumvent censorship.  The censor can rent hosts outside of its own network, but otherwise has no power to monitor or control traffic outside its borders.  Finally, we assume that the censor has sufficient resources to launch successful insider attacks, and thus is aware of the same details of the  circumvention system as are known to ordinary users.
%We further assume that the censor is unable to subvert commonly accepted cryptographic systems, such as 128-bit AES and SHA-256.

Similar to many existing systems~\cite{tor, infranet, collage, Karlin2011, Wustrow2011,Houmansadr2011a}, our approach requires that users run specialized circumvention software on their computers.  We assume that users are able to obtain authentic copies of the software without alerting the government to this fact through some form of out-of-band communication.  (We acknowledge, however, that secure and reliable mechanisms for distributing such software are an important area of future research.)
%We study how to reliably disseminate the bootstrapping materials in our future work.  

\subsection{System Goals}
\label{ssec:goal}

CensorSpoofer aims to achieve the following goals:

{\it Unblockability}: The censor should not be able to block CensorSpoofer without incurring unacceptable costs. %In other words, CensorSpoofer should stay unblockable unless the censor prohibits certain legitimate Internet application (such as IM, Email, and VoIP) entirely.

{\it Unobservability}: The censor should not be able to tell whether a user is using CensorSpoofer or not. 

{\it Perfect resistance to insider attack}: The censor should not be able to break unblockability or unobservability of CensorSpoofer even if nearly all users are corrupted.

{\it Low latency}: CensorSpoofer should be able to provide low-latency communication, such as web browsing, with acceptable quality of service.

{\it Deployability}: CensorSpoofer should be deployable by people with limited resources, without requiring any support from network infrastructure.

%\subsection{Design}
%\label{ssec:design}

%he intuition of our design is as follows. Since the censor can monitor all the traffic of a client, the client cannot connect to a blocked website directly. Hence, the client needs a {\it helper} outside the censor's network to fetch the webpage and send it back to the client. Note that the helper is reluctant to reveal its IP address to the client, e.g. by exposing its IP in the source IP field of the packet header, because the client may be corrupted and the censor could block the helper with source IP filtering. 

%Therefore, we adopt a natural strategy to hide the helper's address from corrupted users, by letting 

%To conceal the helper's IP address, we let the helper spoof another host's IP when sending the censored data back to the client (we refer to this host as {\it dummy destination} and the helper as {\it spoofer}, as shown in Figure~\ref{fig:framework}). To avoid deep packet inspection based censorship, the traffic should be encrypted Meanwhile, we let the client send some dummy traffic to the dummy destination, so that the censor will believe that the client is communicating with the dummy destination legitimately. There are several questions to be answered to make this strategy work against smart censors. %We discuss several of them now, and cover those very detailed issues in Section~\ref{sec:design}.

\section{CensorSpoofer Framework}
\label{sec:framework}

\subsection{Overview}
\label{ssec:intuition}

%\begin{comment}

\begin{figure*}[t]
	\centering
	\includegraphics[height=5.5cm]{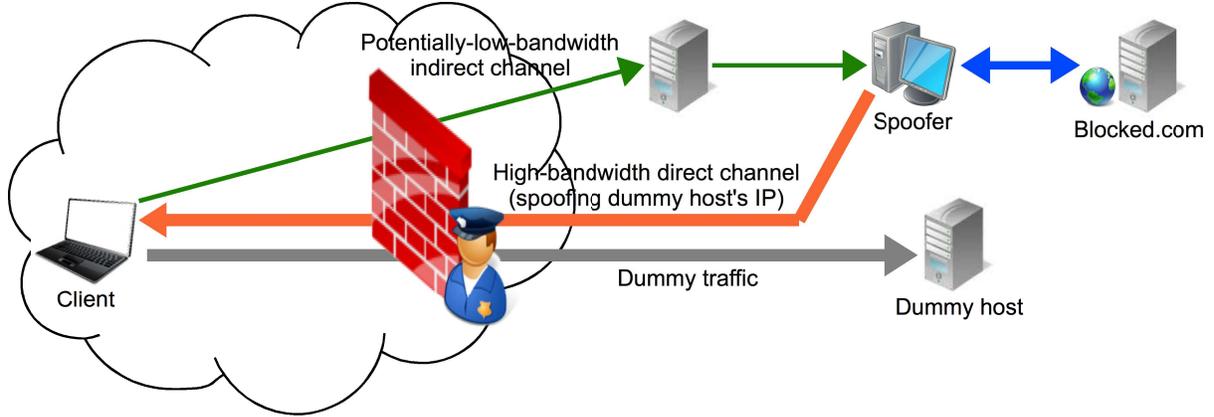}
	\caption{The CensorSpoofer framework. The user pretends to communicate with an external dummy host legitimately, and sends URLs to the spoofer via a low-bandwidth indirect channel (e.g., steganographic IM/email). The spoofer fetches blocked webpages according to the received URLs, and injects censored data into the downstream flow towards the user by spoofing the dummy host's IP. }
	\label{fig:framework}
\end{figure*}

%\end{comment}

In censored countries, users cannot visit blocked websites directly and have to connect to some external relays to access these websites. These relays' IP addresses are exposed to users who connect to them, and therefore can be easily blocked by the censor who colludes with corrupted users.
A natural solution to this is to employ indirect channels to hide the relay' IP. 
%for the communication between the relay node and the user, so that the relay node's IP can be preserved from users and meanwhile blocked webpages can be delivered to the user with acceptable latency. 
For example, MailMyWeb~\cite{mailmyweb} and FOE~\cite{foe} use email as the indirect channel for which the intermediate nodes are Email servers.

To carry voluminous downstream traffic (e.g., web content), the indirect channel must have {\it high bandwidth}. This requirement excludes steganographic indirect channels, such as steganographic IM/email. As a result, the circumvention system has to rely on an encrypted indirect channel so as to utilize full capacity of the indirect channel while avoiding content-based blocking. This requires the intermediate nodes of the indirect channel to support encryption (e.g., TLS/SSL) and reside outside the censor's network (to avoid eavesdropping by corrupted intermediate nodes). 
Currently, only a few email providers can meet this requirement: Gmail, Hotmail, and Yahoo! Mail. However, due to their limited user base in the censored country, the censor could simply block them altogether, as witness when Gmail was blocked in China in 2011~\cite{gmail_block}.

{\bf Our insights.} We notice that for web browsing, the outbound traffic (e.g., URLs) is much lighter-weight than the inbound traffic. If an indirect channel is only used to send outbound messages, high bandwidth is no longer required for the indirect channel. This allows us to use {\it any} indirect channel with steganography to transmit outbound data.  Besides, by using steganography, users can even use local IM or email providers that potentially collude with the censor to access our circumvention system without being detected. The elimination of requiring special servers to provide the indirect channel makes it substantially harder for the censor to block our circumvention system as all overseas Email and IM communication has to be prohibited.

As for the inbound channel, since the relay's IP (i.e., source IP) is not used in packet routing, we can adopt IP spoofing to conceal the relay's IP address. This eliminates the need for an indirect channel to hide the relay's IP, allowing us to use direct channels, which are more common and higher-bandwidth, to send inbound traffic.

{\bf Our design.} Based on these insights, we design a new circumvention framework for web browsing, which uses asymmetric communication with separate inbound/outbound channels. 
In particular,  a user who requires circumvention service first starts or pretends to start a legitimate communication session (e.g., a VoIP call) with a {\it dummy host} residing outside the censor's network, and the relay (called {\it spoofer}) injects censored data into the downstream flow sent to the user by spoofing the dummy host's IP, so that the censor believes the user is legitimately communicating with the dummy host {\it only}. The dummy host does not need to actively cooperate with the user or the spoofer, but should look legitimate to the censor, e.g., its port for VoIP should be open if the cover session is a VoIP call. Meanwhile, the user sends outbound messages containing URLs to the spoofer through a low-bandwidth indirect channel, such as steganographic IM/Email. An illustration of the framework is provided in Figure~\ref{fig:framework}. 

Next, we discuss the inbound and outbound channels in more details.

\subsection{Inbound Channel}
\label{ssec:downstream}   

1) To conceal the spoofer's IP address, we apply IP spoofing in the downstream flow. Then, the first question is {\it what kind of traffic (TCP or UDP) is suitable for IP spoofing?}

Generally, hijacking TCP with IP spoofing is difficult. In TCP, end hosts maintain connection state and acknowledge received data. Suppose the client has established a TCP connection with the dummy host, and the spoofer knows the dummy host's IP address and sequence number and tries to inject packets containing censored data into the downstream flow. First of all, the TCP connection with the dummy host must be kept alive; otherwise, the dummy host will send RST packets in response to the client's packets, which can be easily detected by the censor. In addition, if the spoofer sends more data to the client than the dummy host (i.e., the sequence number of the spoofer is higher than that of the dummy host), the censor can detect the inconsistency of the sequence numbers as long as the dummy host sends any packet to the client\footnote{An active censor can check the dummy host's current sequence number by replaying a client's packet that is outside the dummy host's receiving window; in this case the dummy host will reply an ACK packet containing its current sequence number.}. Thus, the spoofer has to use the sequence numbers that have already been used by the dummy host (i.e., injecting packets as ``resent packets''). However, in this case a censor with packet-recording capability can detect the injected packets by comparing the contents of packets with the same sequence number. 
%Therefore, hijacking TCP with IP spoofing is more suitable for a passive censor, for which the spoofer can take over the downstream channel by closing the connection after TCP handshake.

In contrast, UDP is a connectionless protocol and easier to hijack. Unlike TCP, end hosts of UDP do not maintain any connection state or acknowledge received data. Hence, if the dummy host remains ``quiet'' and the client and the spoofer cooperate closely by sharing initial information and following a proper traffic pattern, it is feasible to deceive a smart censor into believing that the client is legitimately communicating with the dummy host over a duplex UDP channel. In this work, we focus on UDP traffic for IP spoofing. We present a concrete example of hijacking UDP in Section~\ref{sec:design}. 

2) To ensure unobservability, the communication between the client and the spoofer (and the dummy host) should look like a normal UDP session of a legitimate Internet application. So, the second question is {\it what carrier applications should be used?}

UDP is mainly used for time-sensitive applications, such as VoIP, video conferencing, multi-player online games, webcam chat, online TV, etc. These applications usually have high-bandwidth channels. Some other UDP applications, such as DNS and SNMP, have very limited bandwidth and thus are not suitable to carry voluminous downstream traffic.  

We can further divide these applications into two classes based on their communication manner: (1) client-to-server communication, e.g., multi-players online games and online TV, and (2) client-to-client communication, e.g., VoIP and video conferencing. To achieve better robustness to blocking, we prefer the applications in the second class, since for these applications the pool of dummy hosts is significantly larger (e.g., the dummy hosts could be any VoIP client on the Internet), making it much harder to block them altogether.

%To ensure unobservability, censored data must be invisible to the censor. The invisibility can be provided by applying encryption to the censored data. For example, VoIP clients can use encryption protocols, such as SRTP~\cite{srtp} and ZRTP~\cite{zrtp}, to prevent eavesdropping. Due to its client-to-client communication, the VoIP systems can implement encryption only at end hosts, and do not require VoIP servers to support encryption. Therefore, the users only need to install a client software that supports SRTP and/or ZRTP, and can choose any VoIP provider, including local providers controlled by the censor, to access our circumvention system. Currently, there are many VoIP client softwares supporting SRTP and/or ZRTP, such as Blink~\cite{blink}, SFLphone~\cite{sflphone}, Zfone~\cite{zfone}, PJSUA~\cite{pjsua}, etc. 

3) In CensorSpoofer, we use a dummy host as a cover to stealthily transmit censored data. The third question is {\it how to select dummy hosts?} 

%Furthermore, since the IP addresses of dummy hosts are known to users (and further to the censor), we should ensure that the censor is unwilling to block them entirely. 
The selection of dummy hosts is decided by the carrier application. For example, if the carrier application is VoIP, then each dummy host should be a potential VoIP client. Note that an active censor can use port scanning (e.g., using \texttt{nmap}~\cite{nmap}) to check if a dummy host is actually running the application, i.e., listening on a particular port (e.g., port 5060 for SIP-based VoIP). In response, we can use port scanning as well to obtain the list of dummy hosts. According to our experience, a dummy host is ``quiet'' (i.e., not sending any reply packet) to incoming UDP packets sent to a specific port, as long as this port is not ``closed'' on the dummy host. In many cases, port scanning is unable to determine whether a particular application is running on a target machine,  since the target machine could be behind a firewall that is configured to filter probe packets. For example, \texttt{nmap} returns ``open$|$filtered'' or ``closed$|$filtered'' when it cannot tell whether the port is open/closed or the probe is filtered.  This ambiguity plays in our favor as it makes a larger number of hosts appear to 
be plausible VoIP endpoints.

%3) An active censor can use port scanning (e.g., using \texttt{nmap}~\cite{nmap}) to check if a dummy host is actually running the application, i.e., listening on a particular port, e.g., port 5060 for SIP-based VoIP. In response, we can use port scanning as well to get the list of dummy hosts. According to our experience, a dummy host is ``quiet'' (i.e., not sending any reply packet) to incoming UDP packets sent to a specific port, as long as this port is not ``closed'' on the dummy host. In many cases, port scanning is unable to determine whether a particular application is running on a target machine,  since the target machine could be behind a firewall that is configured to filter probe packets. For example, nmap outputs ``open$|$filtered'' or ``closed$|$filtered'' when it cannot tell whether the port is open/closed or the probe is filtered.  This ambiguity makes it even harder for the censor to test the legitimacy of dummy hosts.

4) Finally, we note that not all Internet hosts can launch IP spoofing. Some ASes apply ingress and/or egress filtering to limit IP spoofing. The MIT ANA Spoofer project~\cite{spoofer_mit} has collected a wide range of IP spoofing test results, showing that over 400 ASes (22\%) and 88.7M IPs (15.7\%) can be used to launch IP spoofing. 
Therefore, we need to deploy our spoofer in the ASes where IP spoofing is not prohibited. We can utilize some tools, such as nmap and the spoofing tester developed by the Spoofer project~\cite{spoofer_mit}, to test whether a host can perform IP spoofing.

\subsection{Outbound Channel}
\label{sec:upstream}   

To send outbound requests, we use a steganographic channel embedded in communications such as IM or email.  Note that URLs are typically quite short and can be easily embedded into a small number of messages.  Communication requirements can be further reduced by using a pre-agreed list of censored URLs and sending just the index of the desired site.  Likewise, navigation within a site can use relative link numbering, requesting, e.g., the 3rd link from the front page of \url{www.cnn.com}.  Note that steganography requires the use of a secret encoding key to remain invisible; this process can be made resilient to insider attacks by negotiating a separate key with each user.  Specific steganographic constructions and their security are beyond the scope of this work.  An important challenge that we must address, however, is the possibility that the censor will perform blocking based on the recipient's IM identifier or email address; we discuss a solution in Section~\ref{ssec:circumvention}.

\section{A Design of CensorSpoofer}
\label{sec:design}

In this section, we present a design of CensorSpoofer based on VoIP, although it is possible to build it upon other UDP applications, e.g., video chat. We start with providing background knowledge about VoIP systems.

\subsection{Background of SIP-based VoIP}
\label{ssec:sip}

VoIP is an Internet service that transmits Voice over IP-based networks. It employs session control protocols, such as SIP, MGCP, and H.323, to setup and tear down calls. SIP is one of the most widely-used VoIP signal protocols, because of its light weight. In this work, we focus on SIP-based VoIP systems.

SIP is an application layer protocol. It can run on either UDP or TCP. There are three main elements in SIP systems: user agents, location services, and servers. 

\begin{itemize}

\item {\it User agents} are the end devices in a SIP network. They originate SIP requests to establish media session, and send and receive media. A user agent can be a physical SIP phone or SIP client software running on a computer (also called {\it softphone}). A user agent needs a SIP ID, which is signed up at a SIP provider, in order to make and receive SIP calls.

\item {\it Location service}  is a database that contains information about users, such as SIP IDs, the latest login IP addresses, preferences, etc. Location services generally do not interact directly with user agents. 

\item {\it Servers} are intermediary devices that are located within the SIP network and assist user agents in session establishment. There are two main types of SIP servers: {\it registrar} and {\it proxy}.  A registrar receives SIP registration requests and updates the user agent's information (such as the login IP address) into the location service. A SIP proxy receives SIP requests from a user agent or another proxy and forwards the request to another location.

\end{itemize}

%\subsubsection{SIP Session Setup} 

Here is an example to show how a user (Alice) calls another user (Bob). Suppose Alice has signed up a SIP ID \texttt{alice@atlanta.com} at the SIP provider \texttt{atlanta.com}, and Bob got his SIP ID \texttt{bob@biloxi.com} from \texttt{biloxi.com}, and  Alice knows Bob's SIP ID.

When Bob comes online, he first sends a registration request to the registrar of \texttt{biloxi.com} with its current IP address. 
So does Alice to register herself at the registrar of \texttt{atlanta.com}.

\begin{figure}[t]
	\centering
	\includegraphics[width=8cm]{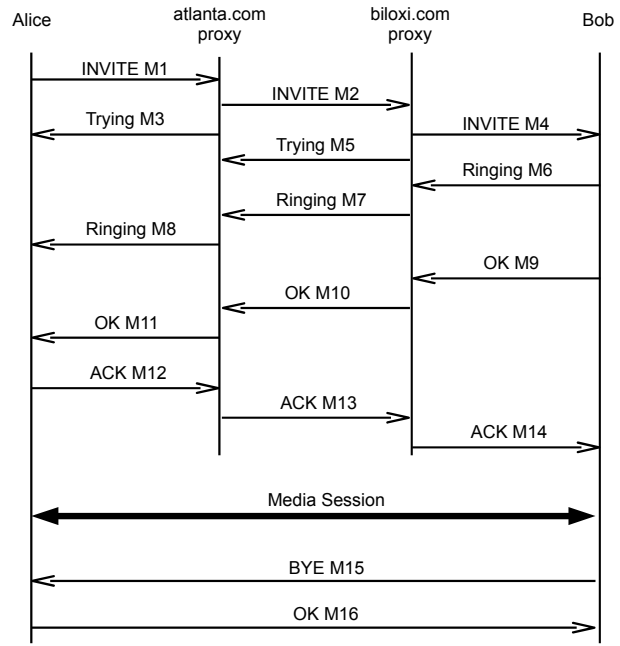}
	\caption{An example of a SIP session (registrars and location services are not shown).}
	\label{fig:sip}
\end{figure}

The SIP call initialization process is shown in Figure~\ref{fig:sip}. First, Alice sends an INVITE message (M1), which contains her SIP ID and IP address, Bob's SIP ID, her supported media codecs, etc., to the proxy of \texttt{atlanta.com} (note that  at this point Alice does not know Bob's IP address). The local proxy performs a DNS lookup to find the IP address of the proxy serving Bob's domain, i.e., \texttt{biloxi.com}, and then forwards the INVITE message (M2) to the remote proxy. At the meantime, the local proxy sends a Trying response (M3) back to Alice, indicating that the INVITE has been received and is being routed to the destination. Upon receiving the INVITE message, the proxy of \texttt{biloxi.com} sends a query to its location service to look up the registered IP address of Bob, and then it forwards the INVITE message (M4) to Bob. The user agent of Bob sends a Ringing response (M6) to the proxy indicating that Bob's phone is ringed. If Bob decides to answer the phone, an OK message containing Bob's current IP (M9) is sent and forwarded back to Alice; otherwise, a Reject response is returned (not shown in the figure). From the received OK message, Alice learns Bob's IP address, and sends an ACK message towards Bob (M12, M13, M14). At this point, the SIP initialization session is done, and Alice and Bob start the media session by sending each other audio data directly. At the end of the media session, either party can send a BYE message (M15) to close the call.

The media session uses Real-time Transport Protocol (RTP) to transmit audio data, and Real-time Transport Control Protocol (RTCP) to provide out-of-band statistic and control information for the RTP flow. Both RTP and RTCP run on top of UDP. 
VoIP clients can use SRTP/SRTCP~\cite{srtp}---an encrypted version of RTP/RTCP---to encrypt their voice communication. SRTP/SRTCP only requires the user to install a user agent that supports RTP/RTCP encryption, and does not require the VoIP servers to support encryption. This implies that the user can use any VoIP provider, including local providers that collude with the censor, to access our circumvention system. Currently, there are many VoIP clients supporting SRTP and/or ZRTP, such as Blink~\cite{blink}, SFLphone~\cite{sflphone}, Zfone~\cite{zfone}, and PJSUA~\cite{pjsua}. The encryption key for SRTP/SRTCP can be either established beforehand, e.g., via MIKEY~\cite{mikey}, or negotiated on the fly using ZRTP~\cite{zrtp}. In this work, we consider using pre-established keys for SRTP/SRTCP.

\subsection{Censorship Circumvention}
\label{ssec:circumvention}

A sketch of the circumvention procedure is as follows. 
The client first initializes a SIP session with the spoofer by sending out a normal INVITE message. Upon receiving the INVITE message, the spoofer randomly selects a dummy host and replies with a manipulated OK message that looks like originating from the dummy host. When the OK message arrives, the client starts to send encrypted RTP/RTCP packets with random content to the dummy host, and the spoofer starts to send encrypted RTP/RTCP packets to the client by spoofing the dummy host's IP address.
Meanwhile, the client sends URLs through a steganographic IM/Email channel to the spoofer. The spoofer fetches the webpages, puts them into RTP packet payloads and sends them to the client.
To terminate the circumvention session, the client sends a termination signal to the spoofer over the outbound channel, and then the spoofer sends a BYE message (with IP spoofing) to the client to close the call.

\subsubsection{Invitation-based Bootstrapping} 

Since the censor can learn the callee SIP ID from the INVITE message, the user cannot use a common callee SIP ID to call the spoofer (otherwise, he/she will be detected once the censor learns the spoofer's SIP ID from corrupted users). There is a similar issue for the steganographic IM/Email channel: the censor can detect users sending IMs or Emails to the spoofer based on the recipient's IM ID or Email address (generally referred to as {\it upstream ID}).

To address this, we let the spoofer use a {\it unique} callee SIP ID and a {\it unique} upstream ID to communicate with each client. 
Hence, the SIP IDs and upstream IDs of the spoofer learned by corrupted users cannot be used to detect honest users. To avoid the bottleneck of having the spoofer create a large number of SIP and upstream IDs by itself, we have each client sign up a callee SIP ID and an upstream ID on behalf of the spoofer, and give them to the spoofer when joining the system. We achieve this by introducing an invitation-based bootstrapping process.

In particular, if a user Alice wants to join the circumvention system, she needs an invitation and help from an existing CensorSpoofer user (say Bob). Alice must trust Bob (e.g., Bob is a friend of Alice); otherwise, Bob could simply report Alice to the censor for attempting to access circumvention service. (We note that similar invitation-based bootstrapping strategies have already been adopted by some real-world circumvention systems, e.g., Psiphon~\cite{psiphon}.)
First, Alice needs to sign up two SIP IDs and two upstream IDs. One pair of SIP ID and upstream ID is for herself, and can be obtained from her local SIP and IM/Email providers which potentially collude with the censor. The other pair is for the spoofer, and must be signed up at abroad SIP and IM/Email providers (not necessarily supporting encryption). If all external SIP, IM, or Email providers are blocked by the censor, Alice can ask Bob to use his already-established circumvention channels to sign up these IDs for her. Then, Alice encrypts the following registration information with the spoofer's public key: \vspace{2mm}

\begin{tabular}{l}
	{\it caller SIP ID $|$ master key $|$} \\
	{\it callee SIP ID $|$ passwd for callee SIP ID $|$} \\
	{\it upstream ID $|$ passwd for upstream ID} 	\vspace{2mm}
\end{tabular}

The master secret is used to derive SRTP/SRTCP session keys (and the key for the steganographic outbound channel if necessary), and the passwords are for the spoofer to login the callee SIP ID and the upstream ID. 

To complete the bootstrapping, Alice needs to deliver the encrypted registration information to the spoofer. Alice could ask Bob to forward the whole registration information to the spoofer through his outbound channel. To reduce the bandwidth consumption of Bob's outbound channel, Alice could let Bob only forward the encrypted upstream SIP ID and password to the spoofer; once her outbound channel is established, she can send the rest registration information to the spoofer by herself.

Note that our unique-ID-assignment strategy cannot be applied to existing relay-based circumvention systems, such as Tor, to improve the robustness against the insider attack. This is because the ``ID'' in CensorSpoofer is an application-level ID, and it is fairly easy to get a large number of them; whereas, in Tor, the ``ID'' that a user use to communication with the relay is the relay's IP address, and IP address is commonly viewed as a scarce resource and it is hard to get a large number of spare IP addresses.   

For the spoofer, it needs to run multiple SIP IDs and multiple upstream IDs at the same time (possibly with a common service provider). In general, IM/Email servers and SIP registrars do not limit the number of accounts registered from a common IP address, because it is possible that multiple legitimate clients are behind a NAT sharing the same IP address. We did some tests on two real-world VoIP providers \texttt{ekiga.net} and \texttt{mixvoip.com} with 100 different SIP IDs running on one of our lab machines, respectively. It turned out for both providers, all these SIP IDs can be registered and receive calls successfully. We also did tests on Gtalk with 10 different accounts on the same machine and all of them worked properly.

\subsubsection{Manipulating the OK Message}

Once the bootstrapping is done, the client can initialize a circumvention session by calling the spoofer using the previously registered callee SIP ID.
In the SIP protocol, the callee's IP address is written into the OK message (more specifically, the enclosed SDP message~\cite{sdp_ietf}, which is used to negotiate the session format, such as codecs, ports, IP, etc.), and later is used by the caller to send RTP/RTCP packets to the callee. Since the OK message can be eavesdropped by the censor, the spoofer cannot put its real IP  into the OK message. 

For this, we use a trick to hide the spoofer's IP address. According to the IETF standards~\cite{sip_ietf, sdp_ietf}, the SDP messages are not checked by SIP proxies. This means the spoofer can put the dummy host's IP, instead of its own IP, into the OK message, without influencing the OK message being forwarded back to the client. Since the registered IP of the callee SIP ID (kept by the location service of the spoofer's VoIP provider) is unknown to the censor, the manipulated OK message is still plausible to the censor. 
To verify the feasibility of replacing the spoofer's IP address in the OK message in practice, we utilized \texttt{netfilter\_queue}~\cite{netfilterqueue} to modify the OK message on the fly, and tested it with two VoIP providers \texttt{ekiga.net} and \texttt{mixvoip.com} and an unmodified VoIP softphone PJSUA~\cite{pjsua}. We found all manipulated OK messages were successfully delivered to the client and the client-side softphone started to send RTP/RTCP packets to the replaced IP after receiving the OK message.

\subsubsection{Selection of Dummy Hosts}
\label{sssec:select_dummy}

A SIP client listens on TCP and/or UDP port 5060 for SIP signalling, and the ports for RTP/RTCP are selected randomly on the fly (usually RTP uses an even port and RTCP uses the next higher odd port). To check the legitimacy of a dummy host, the censor could apply port scanning to test if the ports used by VoIP are open on the dummy host. In response, we can also use port scanning to get the list of dummy hosts.  As we mentioned before, in many cases, port scanning can only return an ambiguous result. For \texttt{nmap}~\cite{nmap} (the state-of-the-art port scanning tool),  the possible probing results include ``open'', ``closed'', ``filtered'', ``unfiltered'', ``open$|$filtered'', ``closed$|$filtered", and ``host seems down''. Only ``closed'' can clearly tell the censor a particular application is not running on the target machine. When the status is ``host seems down'', it is very likely that the target host is offline. For safety, we exclude ``host seems down'' from the acceptable probing states. Therefore, we let the spoofer periodically run port scanning with randomly selected IPs outside the censor's network to get a list of acceptable $\langle ip, rtp\_port\rangle$ (see Algorithm~\ref{alg:port_scan}).

Another strategy for the censor to check legitimacy of the dummy host is to compute the AS path of the spoofing traffic and compare it against the observed entry point of the inbound traffic (i.e., where it enters the censor's network). If the dummy host is located far from the spoofer, it is likely that the entry point of the spoofing traffic is inconsistent with its claimed AS path. To deal with this, we first use \texttt{traceroute} to compute the AS path from the spoofer to the client (called {\it reference AS path}), and then choose a dummy host whose predicted AS path to the client is consistent with the reference AS path with respect to their entry points. Researchers have proposed several AS-path inference algorithms with high predication accuracy (such as ~\cite{QiuPath}).

\begin{algorithm}[t]
%\begin{small}
%\SetLine
\KwIn{$IP\_range$	// outside censored networks}
\KwOut{$dum\_hosts$}

$dum\_hosts \leftarrow \{\}$ \;
$unaccepted \leftarrow \{closed, host\_seems\_down\}$ \;
%$accepted\_res \leftarrow \{open, open|fltr, closed|fltr \} $ \;
\ForEach{$ip \in IP\_range$} {
%	 $res \leftarrow$  port\_scan$(ip, 5060)$ \;
	 \If { \textup{port\_scan}$(ip, sip\_port) \notin unaccepted$ } {
	 		$rtp\_port \leftarrow$ rand\_even\_port() \;
	 		$rtcp\_port \leftarrow rtp\_port + 1$ \;
	 		\If { \textup{port\_scan}$(ip, rtp\_port) \notin unaccepted$ and \textup{port\_scan}$(ip, rtcp\_port) \notin unaccepted$ } {
	 			 add $\langle ip, rtp\_port\rangle$ to $dum\_hosts$ \;
	 		}
	 }
}
%\end{small}
\caption{Port scanning algorithm to find a list of candidate dummy hosts}
\label{alg:port_scan}
\end{algorithm}

In addition, since the port status on a probed host may change over time, we let the spoofer keep track of the previously found dummy hosts and maintain a list of alive dummy hosts. When a circumvention request arrives, the spoofer picks a dummy host from the alive-host list, and keeps checking the VoIP ports of this dummy host during the circumvention session. If the spoofer detects any port of SIP, RTP and RTCP on the dummy host is closed before the circumvention session ends, it sends a BYE message to the client immediately to terminate the SIP session. If the client wants to presume the circumvention session, it needs to initialize another SIP session with the spoofer.

\subsubsection{Traffic Pattern and Bandwidth}

To resist traffic-pattern-analysis attack, the client and the spoofer should follow certain patterns of legitimate VoIP traffic when sending RTP/RTCP packets. For VoIP, both RTP and RTCP packets are of the same size and sent periodically\footnote{Some softphones have the option of Voice Activity Detection (VAD), which can avoid unnecessary coding and transmission of silence voice data. With VAD, the RTP packet size and sending interval may variate. In this work, we assume no VAD is used at the spoofer or the client for simplicity.}. The packet size and sending frequency are defined by the audio codec,  which is negotiated during the SIP initialization session.  
The codec determines the bandwidth of the inbound  channel ($\sim pkt\_size \times freq$). Some codecs that are used to achieve better voice quality can provide higher bandwidth (e.g., 64 Kbps with G.711), while others provide lower bandwidth (e.g., 16 Kbps with iLBC). Note that the same bandwidth is consumed at the dummy host, due to the dummy traffic sent by the client. We can use some bandwidth estimation tools (e.g., \texttt{packet-trains}~\cite{packettrain}) to figure out how much available bandwidth the dummy host has, and based on that, we choose an appropriate codec to avoid consuming too much bandwidth of the dummy host.

\subsubsection{Packet Loss}

UDP does not provide reliable transmission. A RTP packet containing data of a blocked webpage could be lost during transmission, causing failure of reconstructing the webpage at the client. To tolerate packet loss, we can use Forward Error Correction (FEC) codes (e.g., Reed-Solomon code~\cite{RS_code}) inside the inbound channel, so that the client can recover the webpage as long as a certain number of packets are received.

\section{Security Analysis}
\label{sec:security}

We next discuss the security properties of CensorSpoofer against potential passive and active attacks.

\subsection{Geolocation Analysis}

Since the callee's SIP ID and IP address contained in the OK message are transmitted in plaintext,  a sophisticated censor could record all the IP addresses that have been bound to a particular callee SIP ID over time, and try to discover abnormality based on the geolocations of these IPs. For instance, a SIP ID would look suspicious if its registered IPs for two closely conducted SIP sessions are geographically far from each other (e.g., the SIP ID is first registered with an IP in U.S. and 1 hour later it is registered again with another IP in Europe).

To deal with this, instead of picking dummy hosts randomly, the spoofer can choose a set of dummy hosts, which are geographically close, for a particular callee SIP ID, according to an IP-geolocation database (such as~\cite{geo_database}). In particular, for the first-time use of a callee SIP ID, the spoofer randomly selects a {\it primary dummy host }for it, and keeps this information in the user database. For subsequent SIP sessions calling this SIP ID, the spoofer preferentially assigns its primary dummy host for it. If the port status of the primary dummy host becomes ``closed'', the spoofer then preferentially chooses a dummy host from those that have been assigned to this SIP ID (which are also stored in the user database). If none of them is available, the spoofer selects a new dummy host that is geographically close to the primary dummy host for this SIP ID. (Note that the spoofer should make sure that a particular dummy host is not being used by two or more callee SIP IDs at the same time.)

Furthermore, each user can create multiple callee SIP IDs. When a circumvention session is carried out very close to the previous one, or when the spoofer cannot find a suitable dummy host for a callee SIP ID, the user can choose another callee SIP ID instead.

\subsection{User Agent \& Operating System (OS) Fingerprinting} 

The SIP protocol defines the basic formats of SIP messages, but allows user agents (i.e., softphones or SIP phones) to add optional information into the SIP messages, such as the user's display name, timestamps, and the software/hardware information of the user agent. In addition, SIP messages (e.g., INVITE and OK) contain some random identifiers, such as ``To tag'' and ``From tag'', which are generated by the user agent with self-defined length. Additionally, the SIP messages also contain the codecs that are supported by the user agent.

The above information allows a sophisticated censor to fingerprint a particular user agent. As a result, the censor may detect users communicating with the spoofer based on the user-agent fingerprint of the spoofer. To address this, the spoofer can create a number of user-agent profiles based on the popular SIP phones and softphones, and assign one of them to each callee SIP ID. For a SIP session calling a particular SIP ID, the spoofer generates corresponding SIP messages based on the user-agent profile of the callee SIP ID. 

Note that some softphones are only available for certain OSes. For example, SFLphone~\cite{sflphone} can only be used on Linux, and Blink~\cite{blink} is only available for Windows and Mac users. Hence, a sophisticated censor can use OS fingerprinting tools (e.g., the OS detection of \texttt{nmap}~\cite{nmap})  to check if the dummy host's OS is consistent with its user agent (learnt from the user-agent fingerprint). To handle this, the spoofer can also use the OS fingerprinting tool to detect the dummy host's OS, and based on that, assign an appropriate user-agent profile.

\subsection{Traffic Manipulation}

The censor can also try to manipulate traffic flows in order to detect users accessing our circumvention system. 

In anonymous communication systems (e.g., Tor~\cite{tor}), an attacker could use traffic analysis to detect if two relays are on the same path of a flow, by injecting a specific traffic pattern at one relay  (e.g., by delaying certain packets) and detecting the same pattern at the other relay~\cite{watermark_oakland07}. If applying the same attacking philosophy to CensorSpoofer, the censor could delay the packets sent by the user, and detect if there are any changes of the traffic pattern in the downstream flow. However, this attack is based on the precondition that the flows sent and received by the remote host are correlated, and this is not true for VoIP, since each VoIP client sends RTP/RTCP packets periodically, independent of the incoming flow.

Another way to manipulate traffic is to drop packets. Since the spoofer does not actually receive any RTP/RTCP packets from the user, the censor can drop the user's packets without even being noticed by either the spoofer or the user. The softphones and SIP phones can tolerate a small number of random packet loss; but if there are no RTP/RTCP packets received for a certain period of time (e.g., 30 seconds), they will drop the call automatically. Hence, a censor can adopt the following strategy to detect a CensorSpoofer user: it blocks all the RTP/RTCP packets sent to the callee, and checks if the callee  still sends packets to the client after a certain period of time. However, the price of mounting this attack is very high. Since the censor is unable to tell which flow carries censored data, it has to drop all VoIP flows unselectively, causing normal VoIP conversations being interrupted.  

The censor can also alter, reorder, inject or replay RTP/RTCP packets sent to the callee (i.e., the dummy host). However, since a normal VoIP client running the SRTP protocol can simply filter the invalid packets,  such attacks cannot help the censor detect if the callee is a real SIP client or a dummy host.

\subsection{SIP Message Manipulation}

The censor can attempt to manipulate SIP messages. For instance, the censor can manipulate the IP of the callee (i.e., the dummy host) in the OK message, and then check if there are any RTP/RTCP packets sent to the user. Similar to the packet-dropping attack, this attack will make legitimate users unable to make and receive VoIP calls. 

To resist this attack, the spoofer can compute a short keyed hash of the dummy host's IP (and other important data if any) using the SRTP session key, and put the hash value into some random identifiers (e.g., ``To Tag'') in the OK message. The user who knows the session key can use the embedded hash to verify the integrity of the dummy host's IP. If the user detects the OK message is manipulated, it will terminate the SIP session by not sending an ACK response. %We note that this is not a break of the unblockability property, since the censor has to unselectively interrupt the SIP sessions of all VoIP users.

%\subsubsection{Sophisticated Passive Attacks}

%A very powerful censor might attempt to keep track of a user's online records over time, and try to analyze the collected information to detect abnormal activities. For example, for the circumvention without SIP encryption, the censor may store all the IPs (i.e. dummy hosts' IPs) bound to to an identical SIP ID (i.e. the client's callee SIP ID), and check if the changes of the callee's registered IP address are too frequently or dramatic. To address this, we can let each client use multiple callee SIP IDs and also let the spoofer keeps the last assigned dummy host IP for each callee SIP ID and preferably assign the same IP or a geographically close IP to the callee SIP ID for the future sessions.   

%In addition, the censor may also attempt to analyze the concurrent activities of a user in the upstream communication (IM or Email) and the downstream communication (VoIP). For example, it is less common for people to use IM and send Emails while talking on the computer. However, this kind of attack would be well beyond the level of sophistication observed in current censors \pcomment{(use the citation of Telex for this)}.  

%\subsubsection{Denial of Service (DoS)}

\section{Prototype and Evaluation}
\label{sec:experiment}

In this section, we briefly describe our prototype implementation, and report the evaluation results. We refer interested readers to Appendix I for detailed description of our prototype implementation.

\begin{figure*}[t]
\centering
\subfigure[Time to download the html file only. ]{
\includegraphics[height=5.5cm]{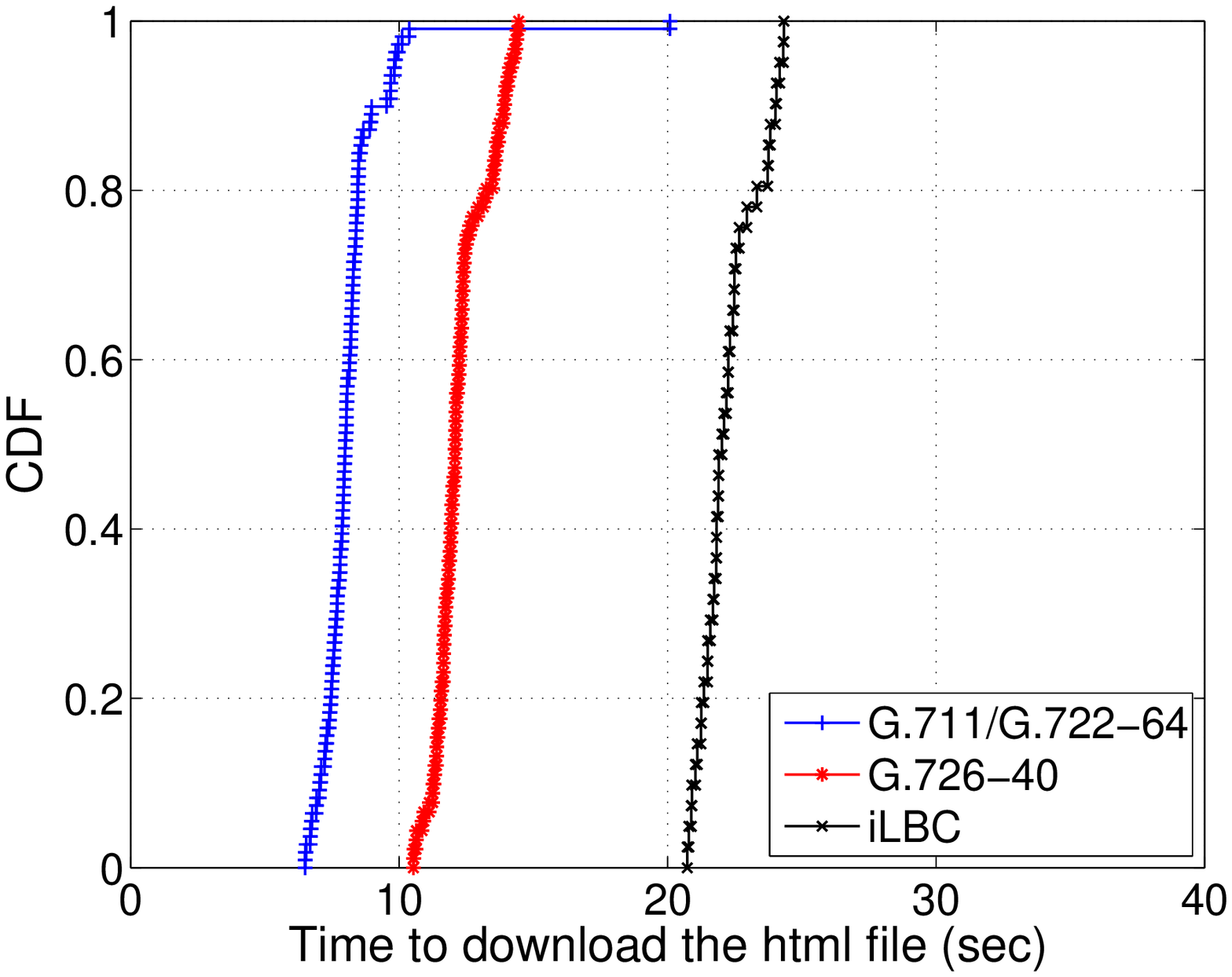}
\label{fig:html}
} 
\hspace{10mm}
\subfigure[Time to download the full webpage.]{
\includegraphics[height=5.5cm]{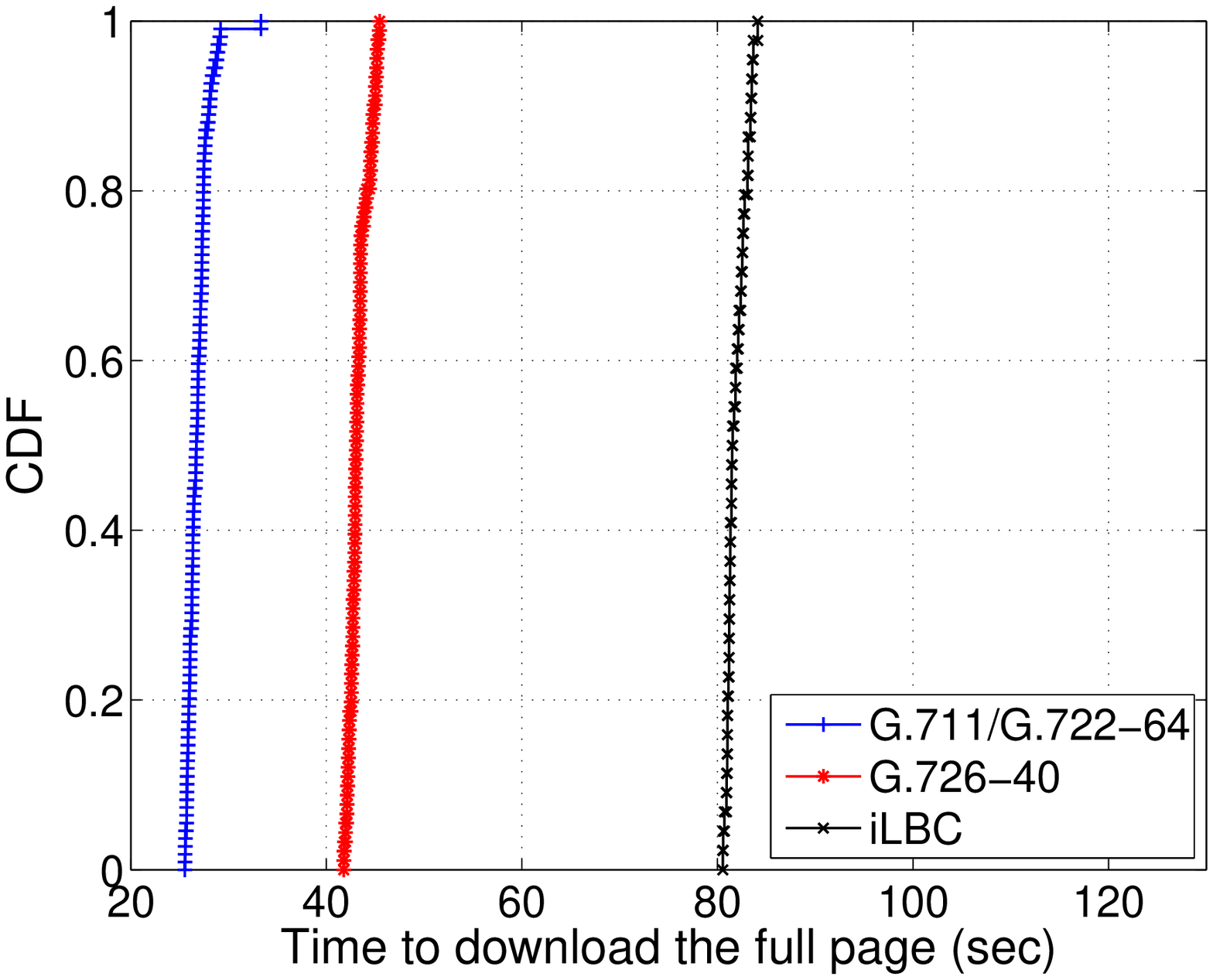}
\label{fig:fullpage}
} 
\caption[]{Performance evaluation. }
\label{fig:performance}
\end{figure*}

\begin{figure*}[t]
\centering
\subfigure[Time to download the html file only. ]{
\includegraphics[height=5.5cm]{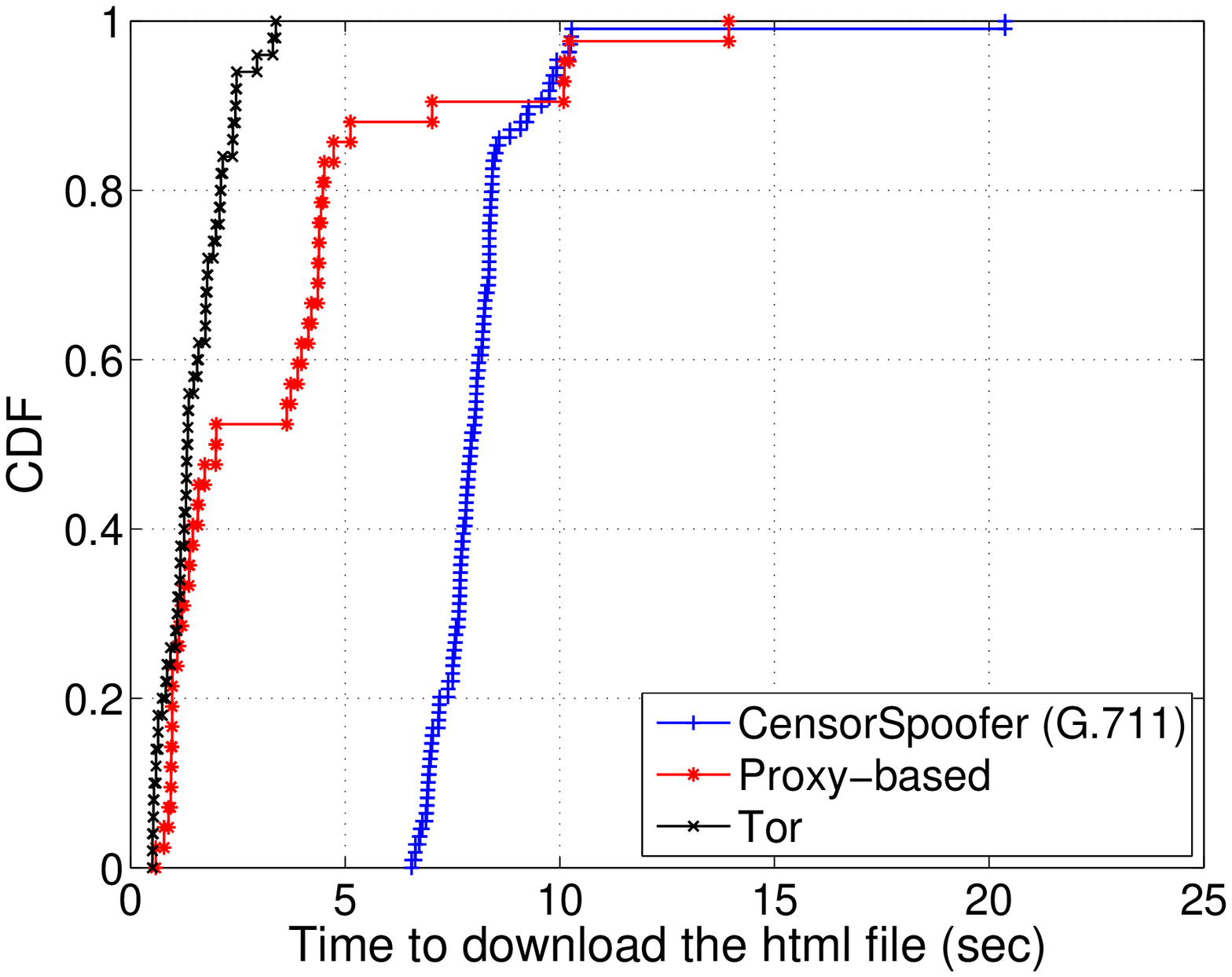}
\label{fig:html}
} 
\hspace{10mm}
\subfigure[Time to download the full webpage.]{
\includegraphics[height=5.5cm]{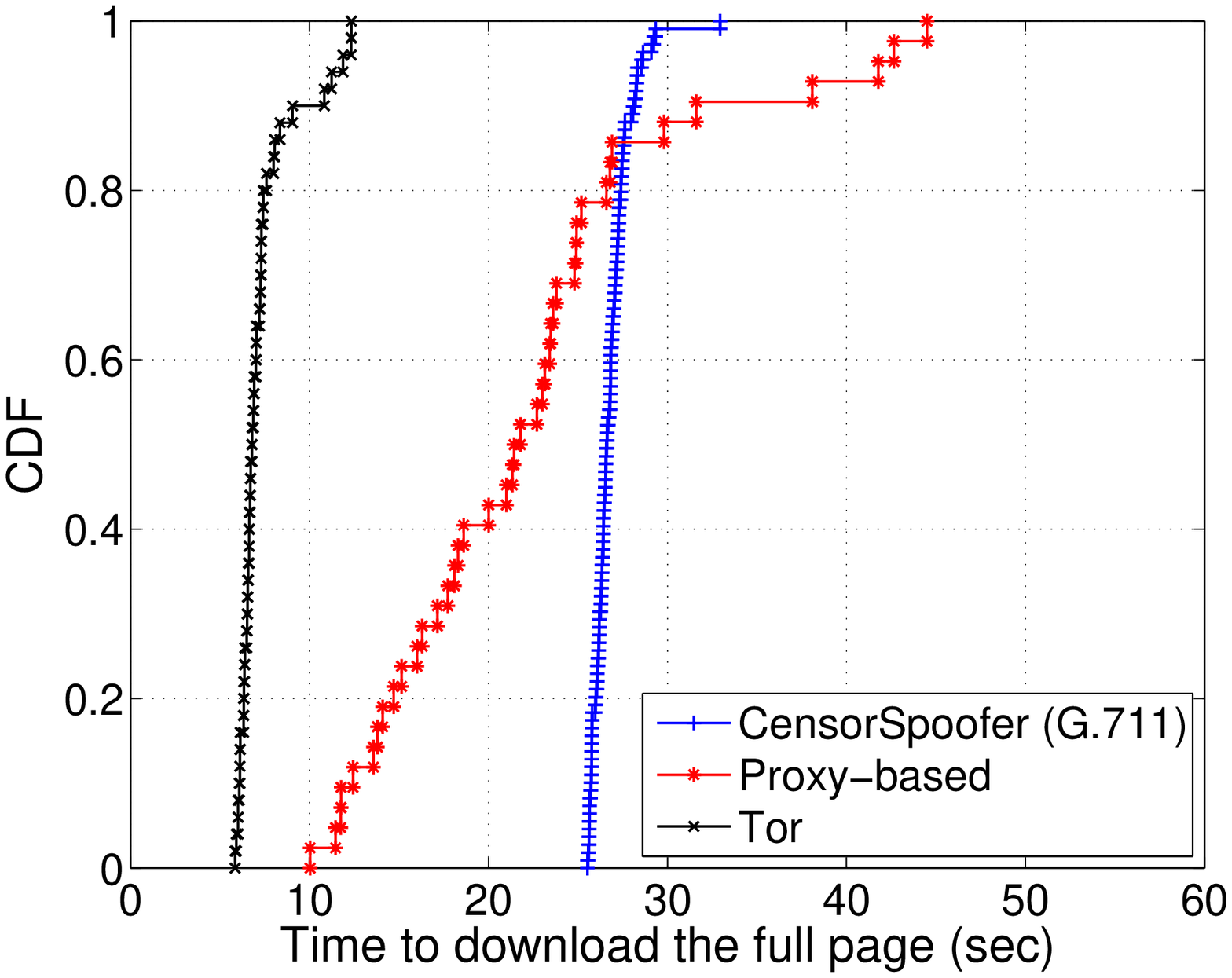}
\label{fig:fullpage}
} 
\caption[]{Performance comparison. }
\label{fig:comparison}
\end{figure*}

\begin{figure*}[t]
\centering
\subfigure[CDF (Note that the CDF plot is truncated to max\_len\_stay\_usable = 6 hours, since many dummy hosts stay usable for a very long time).]{
\includegraphics[height=5.5cm]{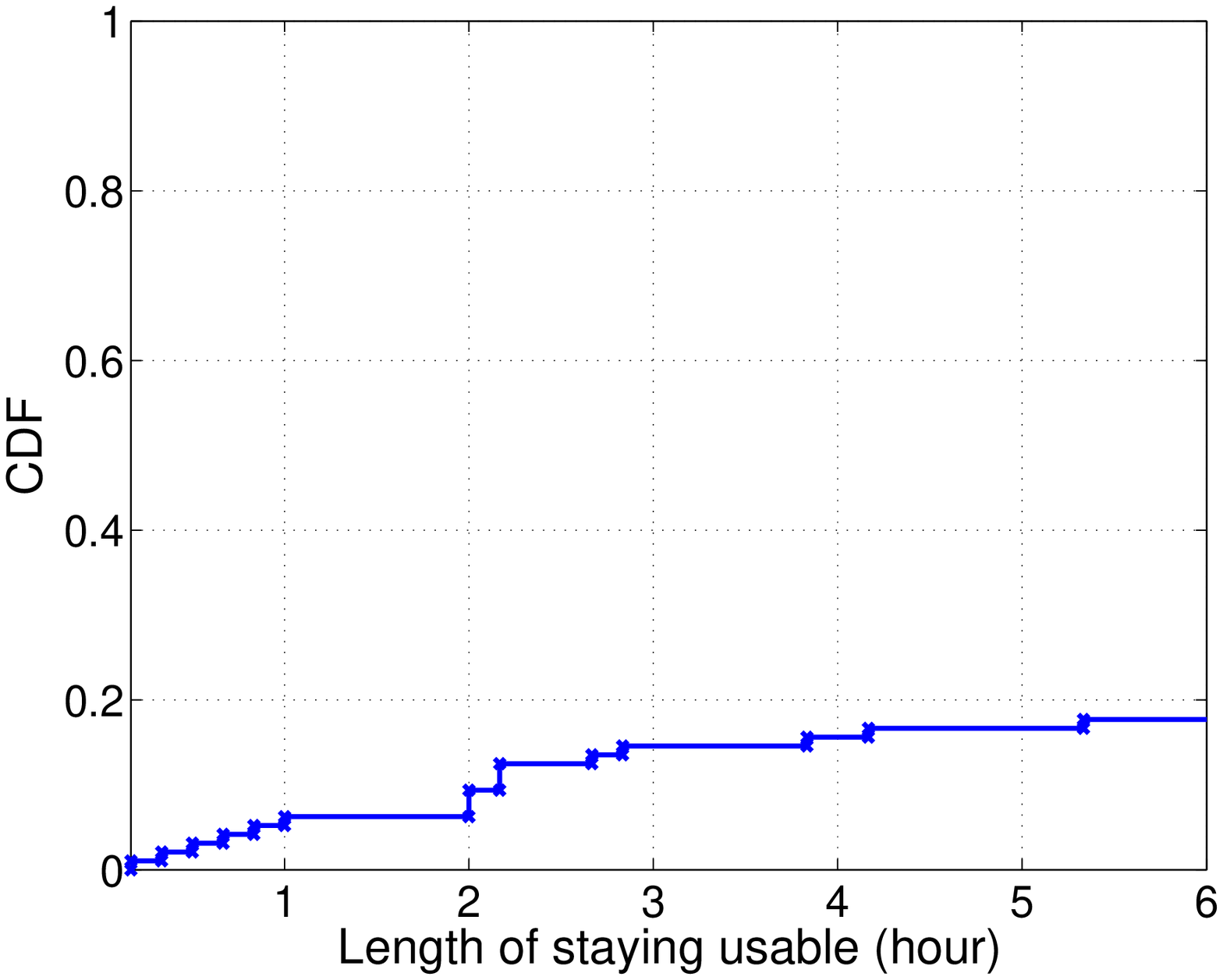}
\label{fig:stab_short}
} 
\hspace{10mm}
\subfigure[Stability of dummy hosts over a long period of time.]{
\includegraphics[height=5.5cm]{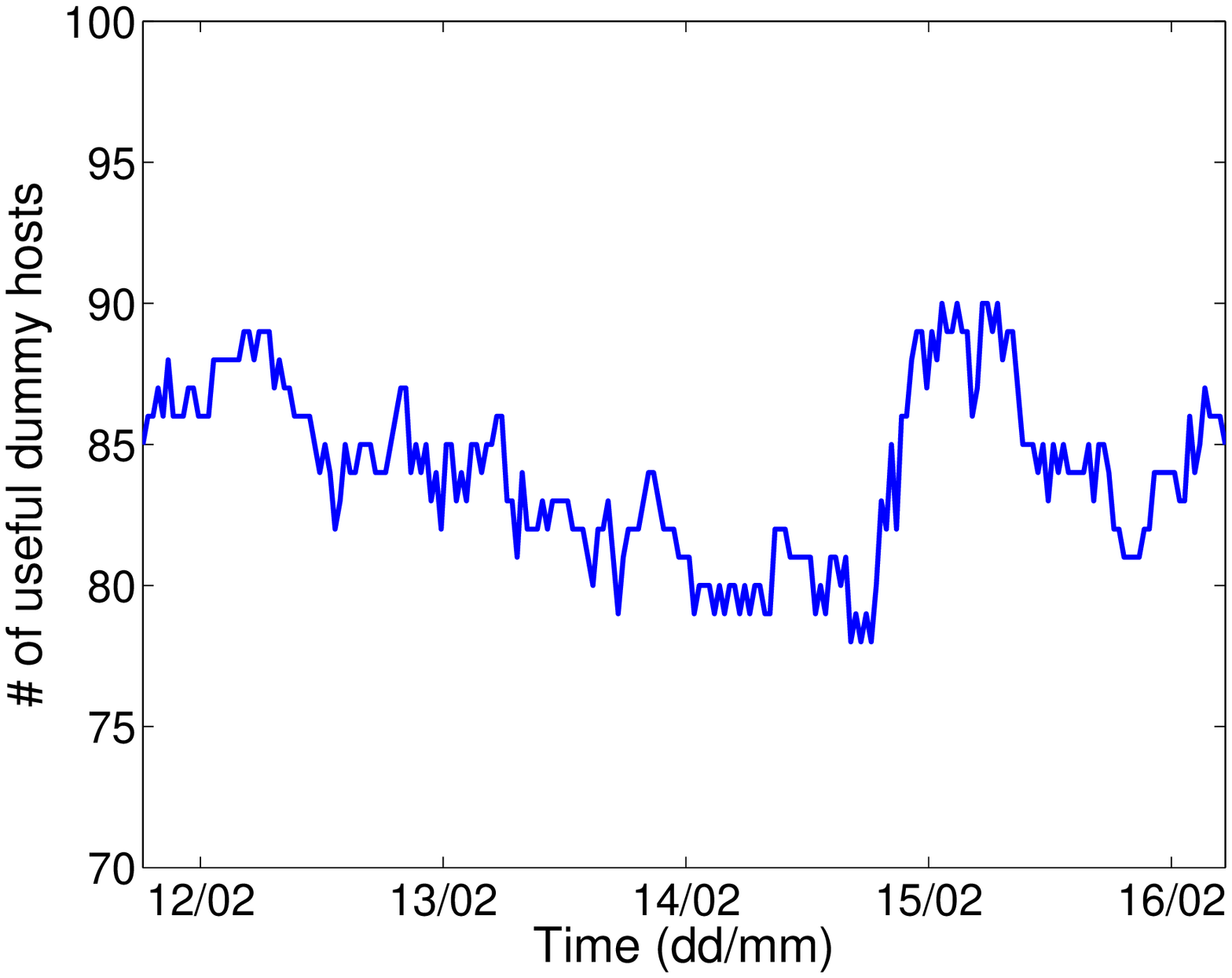}
\label{fig:stab_long}
} 
\caption[]{Stability of dummy hosts. }
\end{figure*}

\subsection{Sketch of Prototype Implementation}
\label{ssec:sketch_prototype}

The spoofer prototype has four components: a SIP message handler, a RTP/RTCP transmitter, an outbound message receiver, and a prefetching proxy. For the SIP message handler, we used \texttt{tcpdump} to create user-agent profiles, and \texttt{netfilter\_queue}~\cite{netfilterqueue} to capture incoming INVITE messages. We used UDP raw sockets to send RTP/RTCP packets. The raw socket allows us to put an arbitrary IP into the source IP field in the IP header. We implemented a XOR-based encoder/decoder to handle packet loss. For this prototype, we used Gtalk as the outbound channel, although our system in no way depends on encrypted indirect channels like Gtalk. We implemented a simple Gtalk client using a python API \texttt{xmpppy}~\cite{xmpppy} to send and receive outbound messages. For ordinary web browsing, a user's web browser sends separate HTTP requests for the html file of the webpage as well as the objects embedded in the webpage. To minimize the number of messages sent through the outbound channel, we implemented a prefetching proxy for the spoofer, which can parse the html file to figure out the missing objects and fetch these objects on behalf of the client, so that the client only needs to send a single HTTP request to the spoofer to download a webpage. Our implementation was based on an open-source layout engine \texttt{QtWebKit}~\cite{qtwebkit}.

As for the client, we implemented a client-side HTTP proxy to handle the HTTP requests made by the user's browser and the HTTP responses received from the RTP channel. The proxy only forwards the first HTTP request to the spoofer via the Gtalk channel and caches the HTTP request-response pairs in memory; when the browser makes a HTTP request, the proxy will serve the browser the appropriate HTTP responses from the memory.  We implemented a minimal browser application -- simply a wrapper around \texttt{QtWebPage} -- to load the webpages and provide statistic information for evaluation.

\subsection{Evaluation}
\label{ssec:evaluation}

We evaluate the performance of CensorSpoofer in a realistic environment, and compare it with other circumvention systems. Then, we measure the selection of dummy hosts.

\begin{table}[t]
\centering
\caption{Bandwidths for different VoIP codecs.}
\label{tab:codec}

%\begin{small}

\begin{tabular}{ | c | c | c |}
\hline
\multirow{2}{*}{Codec}			&	BW of inbound 	& Consumed BW of \\
 											& channel (Kbps)		& dummy host (Kbps)		\\
\hline
G.711								&	64					&		87.2						\\
%\hline
G.722-64								&	64					&		87.2						\\
%\hline			
G.726-40								& 40			& 54.7 \\
%\hline
iLBC		& 		15.6			& 26.6	\\
\hline

\end{tabular}

%\end{small}

\end{table}

\begin{table*}[t]
\centering
\caption{Usable dummy hosts based on AS paths (Spoofer-ASN = 38).}
\label{tab:AS_path}

%\begin{small}

\begin{tabular}{ | c | c | c | c | c | }
\hline
DST-ASN	& \% of direct IPs	& Entry-ASN	&	\# of usable dummy hosts	&	\% of usable dummy hosts	\\

\hline
4134	&	39.4\%	& 4134	& 225	& 100\%			\\
%\hline
4837	&	19.8\%	& 4839	& 225	& 100\%			\\
%\hline			
9394	& 	8.3\%	& 9394	& 217 	& 96.4\%		\\
%\hline
4538	&	7.1\% 	& 23911	& 41	& 18.2\%		\\
\hline

\end{tabular}

%\end{small}

\end{table*}

\subsubsection{Performance Evaluation}
\label{sssec:setup}

The spoofer was deployed on an \texttt{Emulab} machine (located in Utah, U.S.), which has 3.0 GHz 64-bit Duel Core CPU with 1 GB cache and 2 GB of RAM and runs Ubuntu 11. We deployed 8 clients on \texttt{Planetlab}, which are all located in China. Since we aim to evaluate the performance of our system, we let the clients share the same dummy host, which was randomly selected and located in Illinois, U.S. 

To handle packet loss, we made the spoofer add a redundant XOR packet for every 10 packets. We chose the most commonly used VoIP codecs G.726-40, G.722-64, G.711, and iLBC, and set the corresponding RTP packet size and sending interval according to the standard specifications in~\cite{codec}. The bandwidth provided by each codec and the consumed bandwidth of the dummy host are provided in Table~\ref{tab:codec}. 

Each client was configured to repeated download the webpage of \url{wikipedia.org} (which is about 160 KB) for 20 times. For each download, we measured the downloading time for the entire webpage and the html file of the webpage. (Note that once the html file is downloaded, the user's web browser will display the basic frame and the text of the webpage, and the user can start reading the text-based content.)
We found that the clients were able to successfully download the page of \url{wikipedia.org} (which was blocked in China) using CensorSpoofer.
The results of downloading times are provided in Figure~\ref{fig:performance}. We can see that with the codec G.711 or G.722-64, the downloading time for the whole page was 27 seconds, but it only took about 6 seconds to load the html file. 

In addition, we compared the performance of CensorSpoofer with that of existing circumvention systems. We installed a Tor client on one of the \texttt{Planetlab} nodes, and made it connect to a bridge in U.S. to download the webpage of \url{wikipedia.org} for 50 times. Additionally, we ran the same experiment by making the client connect to a public proxy of NetShade\footnote{\url{http://www.raynersoftware.com/netshade/}} (a proxy-based circumvention \& anonymity system), which is located in U.S. Figure~\ref{fig:comparison} shows that it did take longer time for CensorSpoofer to download the pages than the other two circumvention systems, but the downloading time for small web contents, such as html files, for CensorSpoofer is still acceptable. 
%Note that the downloading times of CensorSpoofer are less diverse, because (1) the other two systems are based on TCP using ``timeout-and-resend'', while CensorSpoofer is UDP-based and does not resend packets, and (2) the downstream flow of CensorSpoofer is rate-limited (i.e., sending a RTP packet every 20 -- 30 ms), while the others are not.

We note that the performance of CensorSpoofer can be improved by fixing some limitations of our current implementation. For example, our current prototype of the spoofer does not start sending any packet to the client until it has fully received a html file or an object. We believe removing these limitations can reduce the downloading time. Similarly, the current prototype of the client-side proxy does not deliver HTTP data to the client's web browser until the full html file or object is downloaded. The can be provided by pushing received data to the browser instantly. 

In addition, we notice that the main performance bottleneck of CensorSpoofer is the RTP channel that carries the voice data. We believe by using a higher-bandwidth downstream channel, such as video streaming, the performance of CensorSpoofer can be much improved. 

% either using a higher-bandwidth downstream channel, such as video streaming, or fixing some limitations of our current implementation. For example, our current prototype of the spoofer does not start sending any packet to the client until it has fully received a html file or an object, and there is a similar issue for the client-side proxy. We believe removing these limitations can reduce the downloading time.

\subsubsection{Measurement of Dummy-Host Selection}

To evaluate the easiness of finding dummy hosts, we implemented the port scanning algorithm (i.e., Algorithm~\ref{alg:port_scan} in Section~\ref{sssec:select_dummy}) using \texttt{nmap}~\cite{nmap}. We considered China as the censored country. We randomly selected 10\,000 IPs from the entire IP space, which are located outside China, according to an IP-geolocation database~\cite{geo_database}. We finally found 1213 IPs that can meet our requirements, and the percentage of satisfactory IPs is 12.1\%. This indicates that there are a potentially large number of usable dummy hosts on the Internet. 

Furthermore, we computed the percentage of appropriate dummy hosts for a specific client based on their predicted AS paths to the client. We implemented a widely used AS path inference algorithm~\cite{QiuPath} that is based on AS relationships~\cite{GaoPath}. We considered the top four ASes in China in terms of the number of covered direct IPs (according to~\cite{ChinaAS}), and selected a random IP (i.e., the client) from each of the ASes. We randomly picked 225 dummy hosts out of the 1213 candidate dummy hosts, and computed the AS paths between them and the four clients. Then, we compared the output paths with the AS paths from the spoofer to the clients (computed using \texttt{traceroute}), and filtered the dummy hosts with inconsistent entry points. The results are shown in Table~\ref{tab:AS_path}. We can see that for a specific client, there are enough dummy hosts to use, especially for the clients located in large ASes.

In addition, we measured the stability of dummy hosts over time. Ideally, the dummy host should stay ``usable'' (i.e., none of its VoIP ports becomes ``closed'' or ``host seems down'') during the circumvention session, so that the user does not need to re-initialize the SIP session to change dummy hosts. To justify this, we randomly selected 100 dummy hosts out of the 1213 candidate dummy hosts, kept sending RTP packets to each of them and checking the states of their VoIP ports. Figure~\ref{fig:stab_short} depicts the CDF of length of staying usable for a dummy host. We can see that over 90\% dummy hosts can stay usable for more than 2 hours, and over 80\% can stay usable for longer than 6 hours. This means in most cases, the users only need to establish one SIP session throughout their web browsing. %, which is sufficient for most people's daily web browsing. 

%It is possible that some enterprise-network administrators who see the one-side dummy traffic into their networks would reconfigure their firewalls to block the VoIP ports if these ports are not used, and consequently, the pool of candidate dummy hosts will gradually shrink over time. 

We also measured the stability of dummy hosts over a longer period of time. We kept track of the states of 100 randomly selected dummy hosts from Feb. 9th 2012 to Feb. 16th 2012.
To simulate the practical scenario when the dummy hosts are used by our system to receive VoIP traffic, we kept sending RTP packets to each dummy host periodically, with 1-hour sending period and 1-hour sleeping period. Figure~\ref{fig:stab_long} depicts the number of usable dummy hosts along the time. We can see that the total number of dummy hosts is almost stable, indicating that the overall pool of candidate dummy nodes does not shrink over time.

\section{Conclusion}
\label{sec:conclusion}

In this paper, we proposed a new circumvention framework---CensorSpoofer---that exploits the asymmetric nature of web browsing. CensorSpoofer decouples the upstream and downstream channels, using a
low-bandwidth indirect channel for delivering outbound
requests (URLs) and a high-bandwidth direct channel
for downloading web content. The upstream channel
hides the request contents using steganographic encoding within email or instant messages, whereas the downstream channel uses IP address spoofing so that the real
address of the proxies is not revealed either to legitimate users or censors. 
Unlike some existing circumvention systems, CensorSpoofer does not require any additional support from network infrastructure, and allows individuals to implement the system only at end hosts. We implemented a proof-of-concept prototype for CensorSpoofer, and evaluated it in a realistic environment. The experimental results showed that CensorSpoofer has reasonable performance for real-world usage.

\section{Acknowledgements}

We are grateful to Joshua Juen for his help with the calculation of AS path prediction. We also thank Shuo Tang for helpful discussion on the implementation of prefetching proxy. 

%\newpage

{\footnotesize \bibliographystyle{acm}
\bibliography{sigproc}}

\newpage

\newcommand{\pagefetcher}{PageFetcher\xspace}

\appendix

\section{Appendix I: Prototype Details}
\label{sec:prototype}

\subsection{The Spoofer}
\label{ssec:spoofer}

Our spoofer prototype is mainly composed of the following components: a SIP message handler, a RTP/RTCP transmitter, an outbound message receiver, and a prefetching proxy. 

\subsubsection{SIP Message Handler}

We use PJSUA v1.12~\cite{pjsua} as an out-of-box tool to register the callee SIP IDs. We choose PJSUA because we can easily register multiple SIP IDs using the \texttt{--config-file} option with different configuration files. To prevent the user-agent fingerprinting attack, we use tcpdump to pre-record the OK response messages generated by different softphones, and use them as templates to generate corresponding OK messages to response to different INVITE messages. In our implementation, we create a profile based on the softphone of Ekiga~\cite{ekiga}. 

When starting the spoofer, the SIP message handler first launches PJSUA to register callee SIP IDs, so that the SIP proxies can forward INVITE messages related with these SIP IDs to the spoofer. We use \texttt{netfilter\_queue}~\cite{netfilterqueue} to capture incoming INVITE messages. (Since {PJSUA} requires to bind port 5060, we do not create a socket bound to port 5060 to receive INVITE messages.) For each received INVITE message,  the SIP message handler generates a corresponding OK message, by extracting the session related information (such as the caller's SIP ID, IP address, tags, etc.) from the INVITE message and putting them and the IP address of a dummy host into the pre-recorded OK message. Once the OK message is sent out, the spoofer creates a thread for the RTP/RTCP transmitter for this client.

\subsubsection{RTP/RTCP Transmitter}

The RTP/RTCP transmitter needs to send RTP and RTCP packets periodically with IP spoofing. For this, we use a UDP raw socket, which allows us to put an arbitrary IP into the source IP field in the IP header. To encrypt RTP/RTCP packets, we use AES-128 of OpenSSL v1.0.0~\cite{openssl} with a pre-shared key. Since the sending frequency of RTCP packets is much lower than that of RTP packets, we only use RTP packets to carry censored data and send RTCP packets with randomly generated payloads. 

To handle packet loss, we implemented a simple XOR-based encoder and decoder. The RTP/RTCP transmitter partitions the flow of each task (i.e., downloading a particular webpage) into fixed-sized data blocks (smaller than the RTP payload), and multiplex the blocks of different tasks of the same client into one stream, which is further divided into groups of size $\lambda$ (e.g., $\lambda = 10$ blocks). 
For each group, the transmitter generates a redundant block by XORing all $\lambda$ blocks in the group, so that any $\lambda$ out of  the $\lambda + 1$ blocks are sufficient to recover the whole group. Whenever a RTP packet needs to be sent, the transmitter checks if there are any available blocks (including XOR blocks) in the buffer for this client. If so, it writes one block into the RTP payload and sends it out; otherwise, the RTP packet is stuffed with random data. 

Note that some blocks may contain data less than their capacity (e.g., the last block of a task), and blocks may arrive at the client in different order than being sent out; besides, the client should be able to differentiate blocks for different tasks. To handle these, we use the first 4 bytes of the RTP payload to carry a block sequence number (2 bytes), a task number (1 byte), and block size (1 byte). These fields are encrypted together with the rest RTP payload. 

\subsubsection{Outbound Message Receiver}

For this prototype, we use Gtalk as the outbound channel, although our system in no way depends on encrypted indirect channels like Gtalk. 
Gtalk employs \texttt{XMPP}~\cite{xmpp} as the transmission protocol. We implemented a simple Gtalk client using a python API xmpppy~\cite{xmpppy} to send and receive Gtalk messages. The Gtalk ID of the spoofer is pre-given to the user. Each Gtalk message contains a URL, the user's IP address, and a task number (also contained in the RTP payload).
The outbound message receiver forwards the received Gtalk message to the prefetching proxy by sending a UDP packet, and then the prefetching proxy will start downloading the webpage according to the URL. 

%\subsubsection{Port Scanner}
%to be added by xun ...

\subsubsection{Prefetching Proxy}

For normal web browsing, a user inputs a URL in its web browser, and the browser will then fetch the html file of the webpage as well as the objects used by the webpage, such as figures and video clips. The browser downloads each object by sending a separate HTTP request. 

Since each CensorSpoofer client only sends one URL (instead of separate HTTP requests) to the spoofer,  the spoofer needs to {prefetch} the whole webpage on the behalf of the client. This means that the spoofer needs to first download the html file of the webpage, parse the html file to figure out the missing objects, and then send separate HTTP requests to fetch these objects, and finally send all the downloaded data to the client over the RTP channel.

We built a prefetching proxy (PFP) for this purpose. Instead of implementing a html parser and fetching embedded objects (which are essentially the
operations of a web browser) from scratch, we use an open-source layout
engine \texttt{QtWebKit}~\cite{qtwebkit}, which is a port of the popular \texttt{WebKit}\footnote{http://www.webkit.org/} layout engine into the
\texttt{Qt} application development framework. We choose \texttt{QtWebKit} because
it provides a simple \texttt{QtWebPage} type that significantly reduces
our development effort. Given a URL to load, a \texttt{QtWebPage} performs all
the necessary network operations, including parsing, Javascript execution, etc., in order to render the webpage.
The PFP obtains all the raw HTTP responses for HTTP requests that the \texttt{QtWebPage} makes. As soon as PFP receives a full HTTP response, it sends the
request-response pair to the client over the RTP channel. When
the \texttt{QtWebPage} finishes loading the entire webpage, the PFP sends an ``End-of-Page'' marker to the client, to inform that there will
be no more request-response pair for this webpage.

There are some limitations with our current PFP implementation. The \texttt{QtWebPage} on
the PFP is a distinct browser instance from the client's browser, so the
HTTP requests it generates are likely different from what the client's
browser generates. This is a certainty in the presence of cookies
because the cookies of the client's browser and all HTTP request headers
are not forwarded to the PFP. Another limitation is that the
current PFP disables Javascript on the \texttt{QtWebPage} because Javascript execution
might generate additional HTTP requests after the page has
``finished'' loading (as notified by the \texttt{QtWebPage}), making it hard for
the PFP to determine when to send the ``End-of-Page'' marker.

%{These two proxies communicate using a custom protocol, whereby
 % the client-side specifies the URL of the page, the server-side
  %fetches instructs the QtWebPage}

\subsection{The Client}
\label{ssec:client}

To avoid the censor detecting CensorSpoofer users based on the fingerprint of their softphones, we do not implement our own softphone for the clients; instead, we let the client use any existing softphone to access CensorSpoofer (i.e., for registration and sending SIP messages). Again, we use PJSUA for the client prototype without special reasons.

When running the client, PJSUA is first launched to register the user's SIP ID. Note that most softphones (including PJSUA) do not support making calls  outside the user interfaces. In order to call the spoofer automatically inside our client program,  we use tcpdump to pre-record the INVITE and ACK messages, and send them during the ongoing SIP initialization session with the spoofer (the ACK message needs to be updated according to the OK message before being sent out).

Once the SIP initialization is done, the client creates a UDP socket to receive RTP/RTCP packets from the spoofer and send RTP/RTCP packets to the dummy host. The client uses the pre-shared key to decrypt received packets and stores the decrypted blocks into a buffer. Once a sufficient number of blocks in a group are received, the client uses the XOR-based decoder to recover the original group.

We implemented a client-side HTTP proxy (CSP) to handle the HTTP requests made by the user's browser and the HTTP responses received from the RTP channel. When the CSP receives the first HTTP request for a page, it forwards the URL of the page to the spoofer via the Gtalk channel, but will not forward subsequent requests for other objects of the page. Instead, the CSP
will ``collect'' in memory the HTTP request-response pairs received from the
spoofer, and will serve to the client's browser the appropriate HTTP
responses from its memory when the browser makes a HTTP
request.

We note that any web browser supporting HTTP proxies, such as Mozilla
Firefox\footnote{http://www.mozilla.com/firefox}, can use the CSP
because the CSP provides an HTTP proxy compliant interface. Therefore,
we do not have to modify existing web browsers or implement a new
one. However, for ease of automating experiments, we implement a
minimal browser application (totalling 150 lines of code) that is
simply a wrapper around \texttt{QtWebPage} to load the webpages. This browser
application also outputs various statistics useful for our evaluation.

	% appendix

\end{document}